\documentclass{mrsej}

\usepackage{tabularx}
\usepackage{multirow}
\usepackage{multicol}

\usepackage{euscript}
\usepackage{epsfig}

\usepackage{graphicx}
\usepackage{rotating}
\usepackage{wrapfig}
\usepackage{bm}

\usepackage{soul} 

\begin{document}

\title
[Influence of $^\mathbf{14}$N hyperfine interaction ...]
{Influence of $^\mathbf{14}$N hyperfine interaction on electron nuclear double resonance of boron vacancy in hexagonal boron nitride
}

\author{G.V.~Mamin$^{1,*}$, E.V. Dmitrieva$^{1}$, F.F.~Murzakhanov$^{1}$, I.N.~Gracheva$^{1}$, V.A.~Soltamov$^{2}$, M.R.~Gafurov$^{1}$}

\affiliation{$^{1}$Institute of Physics, Kazan Federal University, Kremlyovskaya 18, Kazan 420008, Russia}
\affiliation{$^{2}$Ioffe Institute, Polytekhnicheskaya, 26, St.Petersburg 194021, Russia}

\email{*E-mail: george.mamin@kpfu.ru}

\ReceivedRevisedAcceptedPublished{????? ??, ????}{?????, ????}{?????, ????}{?????, ????}

\begin{abstract}
The research focuses on the explanation of a phenomenon observed in the spectra of electron-nuclear resonance (ENDOR) pertaining 
to nitrogen atoms adjacent to the boron vacancy (V$_\mathrm{B}^-$) defect in hexagonal boron nitride (hBN). 
The phenomenon is manifested as a shift of the ENDOR spectrum lines with respect to the nitrogen Larmor frequency. It is hypothesized 
that these shifts are indicative of a substantial hyperfine interaction between the V$_\mathrm{B}^-$ defect and the $^{14}$N nuclei in hBN. 
A calculation utilizing second-order perturbation theory was executed to determine the positions of the ENDOR spectrum lines, 
resulting in the formulation of correction equations. The values obtained from the perturbation theory corrections align 
well with the experimental results. The extent of nuclear state admixture into electron states was found to be around 0.04--0.07\%.

\end{abstract}

\pacs{71.70.-d, 71.70.Jp, 75.10.Dg, 76.30.-v, 76.70.Dx, 76.30.Mi, 73.90.+f}

\keywords{ESR, ENDOR, boron vacancy, hBN}

\makehead{2024}{?}{?}{?????}{G.V.~Mamin, F.F.~Murzakhanov, I.N.~Gracheva et al.}

\maketitle

\section{Introduction}
\begin{figure}[ht]
\centering
  \includegraphics[width=0.7\textwidth]{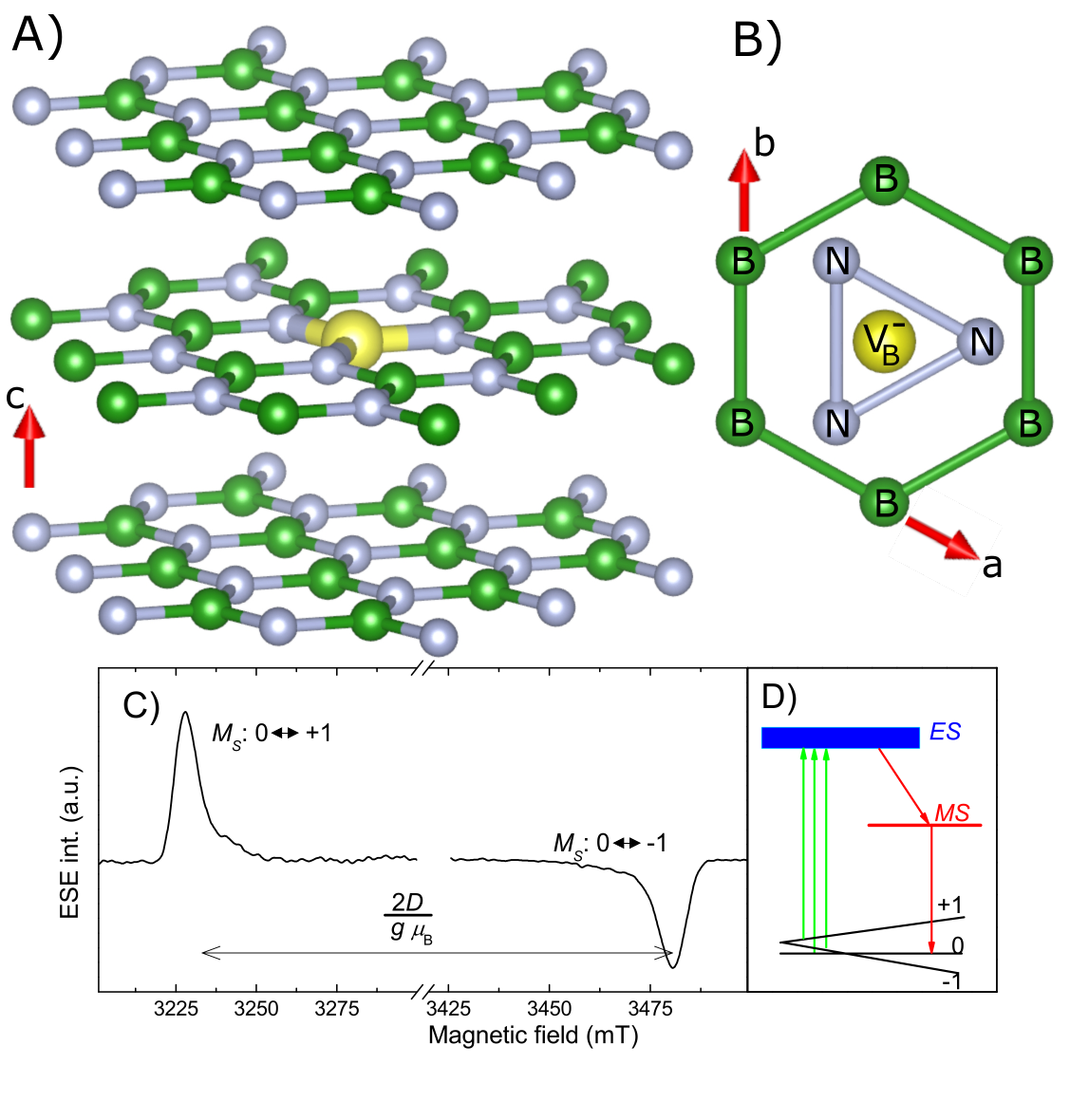}
  \caption{\label{Fig1}
  A), B) Schematic presentation of hBN lattice with V$_\mathrm{B}^-$ defect.
  For B) the $c$-axis is oriented perpendicular to the plane of the figure.
  C) ESR spectrum of the  defects at the temperature of 50 K under 532 nm excitation and  $\mathbf{B_0}\parallel c$. 
  Splitting between fine structure transitions is represented by $\frac{2D}{g\mu_\mathrm{B}}$, 
  with $D$ signifying zero field splitting.
  Fine structure components corresponding to the ESR transition with selection rules $\Delta M_S = 1$ within the $S = 1$ 
  manifold are indicated as $0\leftrightarrow+1$ and $0\leftrightarrow-1$.
  D) Scheme of the  $M_S=0$ spin initialization by 532 nm excitation.}
\end{figure}

Presently, there is an ongoing exploration of quantum technology systems capable of 
functioning without sophisticated and expansive cryogenic facilities 
(at temperatures T~$>$~4.2 K) while accommodating a scalable number of qubits. 
The most recognized spin qubits include the nitrogen-vacancy (NV$^-$) center in 
diamond \cite{l4} and various spin centers found in silicon carbide (SiC) \cite{l2}.
However, the potential for utilizing electron qubits in conjunction with nuclear spin systems 
is constrained to a single nitrogen nucleus unless costly isotopic enrichment with nuclei 
possessing a magnetic moment is employed. In contrast, hexagonal boron nitride (hBN) 
features nuclei ($^{14}$N, $^{11}$B, $^{10}$B) that are each associated with an 
electron center at a boron vacancy (V$_\mathrm{B}^-$), all of which possess a magnetic moment 
(See Figures \ref{Fig1}~A,~B). This characteristic may significantly enhance the application of 
standard CNOT-type operations by leveraging different types of nuclei, thereby 
facilitating the integration of hBN into quantum computing frameworks \cite{cobarrubia2024hexagonal}.
Furthermore, hBN is classified as a two-dimensional (2D) material \cite{liu2022spin}, which enables the utilization of 
paramagnetic centers situated near the surface, particularly through the combination 
of microwave and optical radiation, presenting opportunities for advanced ultrasensitive nanosensors \cite{zhang2020material}.
In contrast, in three-dimensional (3D) materials, the proximity of spin centers 
to the surface tends to degrade their spin coherence time \cite{zhang2017depth}. Additionally, 
the integration of hBN with other 2D materials has the potential to markedly enhance the sensitivity 
of these systems to various external influences.\par
In previous publications from our research group, we demonstrated the efficacy of high-field electron-nuclear double resonance (ENDOR) 
for both qualitative and quantitative analyses of electron-nuclear interactions in (nano)diamonds \cite{yavkin2019epr} and SiC \cite{murzakhanov2024photoinduced}. 
Our investigation into ENDOR has now been broadened to include hexagonal boron nitride (hBN). Through this extension, 
we have identified several novel features that had not been observed in earlier studies. This paper aims 
to elucidate these findings and provide potential explanations for the observed phenomena.

\section{Materials}
The hBN single crystals with dimensions of 900~$\mathrm{\mu}$m$\times$540~$\mathrm{\mu}$m$\times$55~$\mathrm{\mu}$m 
used in this study were commercially produced by the HQ Graphene company. 
The samples were irradiated at room temperature with 2~MeV electrons 
to a total dose 6$\times$$10^{18}$~cm$^2$. No annealing treatments were applied to the irradiated samples.

\section{Methods}
The electron spin resonance (ESR) experiments were carried out on the W-band (operating at the microwave (MW) frequency of 94~GHz) 
Bruker Elexsys E680 commercial spectrometer (Karlsruhe, Germany) can be deleted. 
The ESR spectra in a pulsed mode were registered by detecting the 
amplitude of the primary electron spin echo (ESE) as a function of the magnetic field sweep 
$B_0$ using a pulse sequence $\frac{1}{2}\pi(\mathrm{MW})$--$\tau$--$\pi(\mathrm{MW})$--$\tau$--ESE, 
where $\pi(\mathrm{MW})=80$~ns and $\tau=240$~ns. ENDOR spectra were obtained 
utilizing the Mims pulse sequence\par $\frac{1}{2}\pi(\mathrm{MW})$--$\tau$--$\frac{1}{2}\pi(\mathrm{MW})$--$t_1$--$\pi(RF)$--$t_2$--$\frac{1}{2}\pi(\mathrm{MW})$--$\tau$--ESE
with a 150~W radiofrequency (RF) generator, where $\pi(RF)=160$~$\mathrm{\mu}$s. Low temperature ($T$=50~K)
measurements were conducted by using a flow helium cryostat from Oxford Instruments. 
The sample's photoexcitation was provided by green laser ($\lambda=532$~nm) with an output power of 200~mW.

\section{Results and Discussion}

ESR spectrum of the V$_\mathrm{B}^-$ defects in hBN sample under continuous 
optical excitation and orientation 
of $\mathbf{B_0}$ parallel to the hexagonal $c$ axis is presented in 
Figure~\ref{Fig1}~C along with the scheme of the spin initialization with the light 
(Figure~\ref{Fig1}~D, where ES stands for the excited state, MS is for the metastable state, see reference \cite{liu2022spin}).

For $\mathbf{B_0} \parallel c$ the ESR spectrum can be described by the following spin-Hamiltonian

\begin{equation}\label{eq_hamiltonian}
\begin{split}
&\hat{\mathbf{H}}=g\mu_{\mathrm{B}} B_0 S_z+D \left( S_z^2-\frac{S\left(S+1\right)}{3}\right)+\\
&+\sum_{j=1}^{3}{\left( A_{zz} S_z I_{z(i)}-A_{xx} S_x I_{x(i)}+A_{yy} S_y I_{y(i)}-\gamma\hbar B_0 I_z+P\left(3I_{z(i)}^2-I\left(I+1\right)\right)\right)},
\end{split}
\end{equation}

\noindent where the first two terms determine electronic structure of the spin defect with the $g$-factor ${g=-2.004}$ 
and the zero-field splitting (ZFS) constant $D=3.55$~GHz. The next three terms describe the anisotropic hyperfine interaction (with components $A_{xx}$, $A_{yy}$, $A_{zz}$) of the electron 
magnetic moment with the $^{14}$N nucleus and the last term is nuclear quadrupole interaction (NQI)
 ${P=\frac{1}{h}\frac{e^2Qq}{4I\left(2I-1\right)}=\frac{1}{h}\frac{e^2Qq}{4}}$, where $i$ is the index of the 
 nitrogen nucleus from the three nearest atoms of the immediate environment (Figure \ref{Fig1}) \cite{mackenzie2002multinuclear}. 
 In previous papers \cite{l15} the $C_q=\frac{e^2Qq}{h}=4P$ value has been used.
The electron $S$ and nuclear $I$ spins are equal 1. The interaction between the nuclei is neglected.\par 

The experimental ESR spectrum shown in Figure \ref{Fig1}~C demonstrates a pair of ESR transitions that serve 
as fingerprints of the ZFS structure associated with the $S$ = 1 for the V$_\mathrm{B}^-$ defect, manifesting
itself by the splitting in magnetic field by the value 2$D$ , where $D$ = 3.55~GHz is the zero field splitting of the V$_\mathrm{B}^-$ defect, 
and $\frac{g\mu_B}{\hbar}$=28 GHz/T is its gyromagnetic ratio. 
An important feature of the spectrum is that the fine structure components are inverted in phase relative 
to each other, mimicking a significant deviation of triplet spin sublevels population 
from Boltzmann statistics due to optical excitation and indicating population inversion, 
as shown in the scheme presented in Figure \ref{Fig1}~D.
For the purpose of calculating ENDOR spectra, we take into account only one of the three identical $^{14}$N nuclear spins
In this alignment of the magnetic field relative to the $c$--axis, the hyperfine 
coupling has previously been established to be $A_{xx}\approx 45.5$~MHz \cite{l4,l12}. 
The last two terms in equation~\ref{eq_energy_1st} refer to the interaction of the $^{14}$N nuclear magnetic moment with the static 
external magnetic field and the interaction of $^{14}$N nuclear electric quadrupole moment with 
the electric field gradient produced by the surrounding atoms. The nuclear quadrupole constant, which characterizes the 
the corresponding splitting through nuclear quadrupole interaction (NQI) has been determined to be $C_q\approx 2.11\pm0.08$~MHz \cite{l12}. 
To account for the bond direction \cite{l15}, in the Hamiltonian in crystals coordinates, 
$S_{z}$ must be replaced by $S_{x}$ and $S_{x}$ by $S_{z}$, excluding Zeeman interaction. All three nitrogen nuclei are equivalent for 
$\mathbf{B_0} \parallel c$ axis, and the summation over these nuclei can be removed. Also, we take into 
account the closeness of values of $A_{xx}$ and $A_{yy}$ \cite{l12,l15}, and to facilitate the calculations, we replace $A_{yy}$ by $A_{xx}$. 
Creating new variables: ${Z_e\equiv g \mu_{\mathrm{B}} B_0}$, ${Z_n\equiv \gamma \hbar B_0}$, one can rewrite eq. \ref{eq_hamiltonian} as

\begin{equation}\label{eq_hamiltonian1}
\begin{split}
&\hat{\mathbf{H}}=Z_e+D \left( S_z^2-\frac{S\left(S+1\right)}{3}\right)+A_{xx} S_z I_z-A_{zz} S_x I_x+A_{xx} S_y I_y-Z_n+\\
&+P\left(3I_x^2-I\left(I+1\right)\right).
\end{split}
\end{equation}

In general, to analyse the ESR spectra, it is sufficient to consider the energy 
values obtained while neglecting the off-diagonal elements in the spin-Hamiltonian matrix, and these values 
are presented in equation~\ref{eq_energy_1st}. The corresponding energy level diagram is also shown in Figure \ref{Fig3} 
with ENDOR transitions indicated by arrows. \par

\begin{equation}\label{eq_energy_1st}
\begin{split} 
&E_{1}^{(0)}=- Z_e + \frac{D}{3} +A_{xx}+ Z_n- \frac{3}{2}P, \\
&E_{2}^{(0)}=- Z_e +\frac{D}{3} + P, \\
&E_{3}^{(0)}=- Z_e + \frac{D}{3}- A_{xx}- Z_n -\frac{3}{2} P, \\
&E_{4}^{(0)}=- \frac{2 D}{3}+ Z_n - \frac{3}{2}P, \\
&E_{5}^{(0)}=- \frac{2 D}{3} + P,\\
&E_{6}^{(0)}=- \frac{2 D}{3}- Z_n - \frac{3}{2}P \\
&E_{7}^{(0)}=Z_e + \frac{D}{3} - A_{xx} + Z_n - \frac{3}{2}P, \\
&E_{8}^{(0)}=Z_e +\frac{D}{3} + P, \\
&E_{9}^{(0)}=Z_e + \frac{D}{3} +A_{xx} - Z_n - \frac{3}{2}P,  
\end{split}
\end{equation}

Thus, the following transitions should be observed in the ENDOR spectrum:

\begin{figure}[ht]
\centering
  \includegraphics[width=0.6\textwidth]{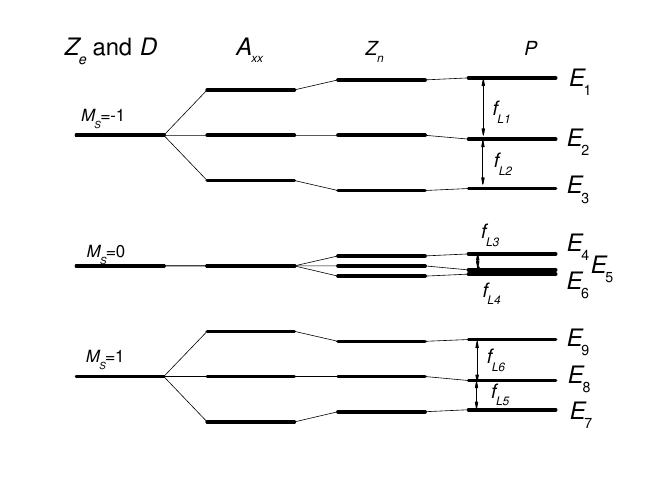}
  \caption{\label{Fig3}
  Schematic diagram of energy levels of the V$_{\mathrm{B}}$-N$_3$ defect.}
\end{figure}

\begin{equation}\label{eq_line_1st}
\begin{split} 
&f_{L1}^{(0)}=A_{xx}+ Z_n + \frac{3}{2} P, \\
&f_{L2}^{(0)}=A_{xx}+ Z_n - \frac{3}{2} P, \\
&f_{L3}^{(0)}= Z_n+ \frac{3}{2} P, \\
&f_{L4}^{(0)}= Z_n- \frac{3}{2} P,\\
&f_{L5}^{(0)}=A_{xx}- Z_n - \frac{3}{2} P, \\
&f_{L6}^{(0)}=A_{xx}- Z_n + \frac{3}{2} P,
\end{split}
\end{equation}
where the lower index $Li$ corresponds to ENDOR transition on Figure~\ref{Fig3}. 
According to equation~\ref{eq_line_1st}, one can identify two groups of symmetrical lines 
(pairs) that have to appear in the ENDOR spectrum. 
The first set, comprising lines $L3$ and $L4$, is found in proximity to the nuclear 
Zeeman value ($Z_n$), and the second quadrupole pair ($L1, L2$ or $L5, L6$) is positioned near the 
$A_{xx} \pm Z_n$ values. The separation between the pairs within each group is equivalent to 3$P$.

\begin{figure}[ht]
\centering
  \includegraphics[width=0.7\textwidth]{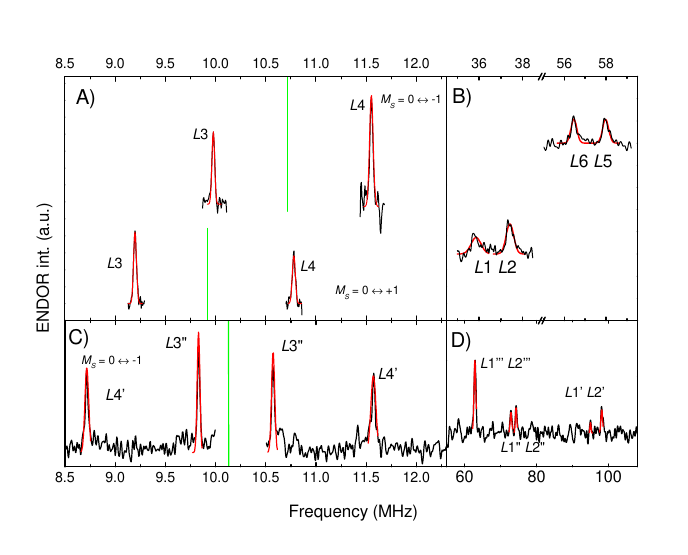}
  \caption{\label{Fig2}
  $^{14}$N ENDOR spectrum for two electronic transitions with 
  $M_S=0\leftrightarrow+1$  ($B_0(L)=3223.1$~mT, top) and $0\leftrightarrow-1$ ($B_0(H)=3480.5$~mT, lower spectrum) recorded at T = 50 K. 
  The numbers ($L1$--$L6$) correspond to the distinct resonance transitions presented 
  in the energy diagram (Figure \ref{Fig3}). $^{14}$N Larmor frequency in a given magnetic fields is indicated by a green lines.
  A, B) $\mathbf{B_0} \parallel c$. C, D) $\mathbf{B_0} \perp c$.}
\end{figure}

The ENDOR spectra  were acquired on both fine structure components low field ($L$): $0\leftrightarrow+1$ (${B_0(L) = 3223.1}$~mT, lower spectrum) and high field ($H$): 
$0\leftrightarrow-1$ (${B_0(H) = 3480.5}$~mT, upper spectrum) 
as shown in Figure~\ref{Fig2}~A,~B for $\mathbf{B_0} \parallel c$ and $0\leftrightarrow-1$ (${B_0(L) = 3287.15}$~mT, Figure~\ref{Fig2}~C,~D) for $\mathbf{B_0} \perp c$ axis. 
For $\mathbf{B_0} \perp c$, the three nitrogen nuclei become nonequivalent (see Figure \ref{Fig1}~B),  and, therefore, three spectra with slightly 
different parameters are observed for each type of nuclei. The corresponding lines are designated by upper indices as ', '', and '''. 
The calculated $Z_n$ values for both values ${B_0(L)}$ and ${B_0(H)}$ are shown in Figures~\ref{Fig2}~A,C by green vertical lines. 
As one can clearly see already from Figure~\ref{Fig2}, the lines in pairs are arranged asymmetrically relative to the $Z_n$.

To obtain more accurate numerical parameters of the ENDOR spectra, the observed ENDOR lines are fitted by Gaussian curves 
(red curves in Figure~\ref{Fig2}). Positions and widths of the resonant nuclear transitions are summarized in Table~\ref{table1}. 
Additionally, the pair spliting of lines in pair and the value of center of the pair (pair midpoint) added to Table~\ref{table1} for comparing with 
NQI spliting and values of $Z_n$ ($B_0(L) = 3223.1$~mT)~=~9.9198~MHz,  $Z_n$($B_0(H) = 3480.5$~mT)~=~10.7120 MHz for $\mathbf{B_0} \parallel c$  and $Z_n$($B_0(L) = 3287.1$~mT)~=~10.1167~MHz for $B_0 \perp c$. 
The ENDOR linewidth $\Delta f$ is used to estimate the error bars in the line position designated as $\frac{\Delta f}{5}$.

\begin{table}[h!]
\begin{center}
\caption{\label{table1}Key parameters of $^{14}$N ENDOR line positions. 
The line splitting corresponds to distance between
 $f_{L3}$ and $f_{L4}$, or $f_{L1}$ and $f_{L2}$, or $f_{L5}$ and $f_{L6}$ frequencies in pairs.
 Midpoints of pairs are calculated as $\frac{f_{L3}+f_{L4}}{2}$, $\frac{f_{L1}+f_{L2}}{2}$, $\frac{f_{L5}+f_{L6}}{2}$ value}
\begin{tabular}{|m{5 cm}|m{0.5 cm}|m{3 cm}|m{2.5 cm}|m{2.5 cm}|}
\hline
             & N  &Frequency,~MHz       & Spliting of lines in pair,~MHz & Midpoint of pair,~MHz \\
\hline
\multirow{4}{*}{$B_0(L)=3223.1$~mT, $B_0 \parallel c$} &  L3 & 9.198$\pm$0. 005   & \multirow{2}{*}{1.58$\pm$0.01} & \multirow{2}{*}{9.99$\pm$0.01}\\
\cline{2-3}
&L4                &  10.778$\pm$0. 005  &                     &              \\
\cline{2-5}
&L1 & 35.9$\pm$0.2 & \multirow{2}{*}{1.6$\pm$0.3} & \multirow{2}{*}{36.65$\pm$0.3}\\
\cline{2-3}
&L2 & 37.4$\pm$0.2 & & \\
\hline 
\multirow{4}{*}{$B_0(H)=3480.5$~mT, $B_0 \parallel c$} & L3 & 9.977$\pm$0. 005   & \multirow{2}{*}{1.58$\pm$0.01} & \multirow{2}{*}{10.76$\pm$0.01}\\
\cline{2-3}
&L4&  11.552$\pm$0. 005  &                     &              \\
\cline{2-5}
&L5 & 56.5$\pm$0.1 & \multirow{2}{*}{1.5$\pm$0.2} & \multirow{2}{*}{57.25$\pm$0.2}\\
\cline{2-3}
&L6 & 58.0$\pm$0.1 & & \\
\hline 
\multirow{4}{*}{$B_0(L)=3287.15$~mT, $B_0 \perp c$ }& L3' & 8.718$\pm$0. 005   & \multirow{2}{*}{2.85$\pm$0.01} & \multirow{2}{*}{10.144$\pm$0.01}\\
\cline{2-3}
&L4'&  11.571$\pm$0. 005  &                     &              \\
\cline{2-5}
&L3" & 9.83$\pm$0. 005 & \multirow{2}{*}{0.7428$\pm$0.01} & \multirow{2}{*}{10.201$\pm$0.01}\\
\cline{2-3}
&L4" & 10.5728$\pm$0. 005 & & \\
\hline 
\end{tabular}
\end{center}
\end{table}

The data presented in Table~\ref{table1} indicates that the midpoint of each pair of lines $L3$ and $L4$ is displaced 
towards higher frequencies relative to the Larmor frequency, specifically by approximately $0.06\pm0.01$~MHz for $\mathbf{B}\parallel c$ 
and $0.03\pm0.01$ and $0.09\pm0.01$~MHz respectively, for $\mathbf{B}\perp c$.

Additionally, the corresponding NQI splitting can be determined as $3P=1.58\pm0.01$~MHz (Table~\ref{table1}) or $P=0.53$~MHz, with the $C_q$ value calculated as $4P=2.11\pm0.01$ MHz 
for crystall orientation $\mathbf{B}\parallel c$.
For $\mathbf{B}\perp c$, the value $3P=$ ranges from 2.58~MHz to 0 consistent with the function $3 cos^2 \theta -1$, where 6$^o$, 54$^o$, and 66$^o$  
represents the angle between main axis for each nonequivalent type of nitrogen nucleus and magnetic field direction. 
Note that this study does not consider the non-axiality of the NQI $\eta$=0.07, resulting in increased
$C_q$ by factor $(1+\eta)=1.07$ time compared to reference \cite{l15}.

In the following analysis, we will explore the effects of the off-diagonal terms of the spin Hamiltonian matrix by applying perturbation theory. 
The diagonal segment of the spin Hamiltonian matrix, as outlined in equation~\ref{eq_energy_1st}, 
will be employed as the initial approximation. As a perturbation operator, we take the lagging off-diagonal part of the spin Hamiltonian matrix:

 \begin{equation}\label{eq_pert_matrix}
\hat{\mathbf{V}}=\left[\begin{matrix}0 & 0 & \frac{3 P}{2} & 0 & - a_{zpx} & 0 & 0 & 0 & 0\\ 
0 & 0 & 0 & - a_{zmx} & 0 & - a_{zpx} & 0 & 0 & 0\\ 
\frac{3 P}{2} & 0 & 0 & 0 & - a_{zmx} & 0 & 0 & 0 & 0\\ 
0 & - a_{zmx} & 0 & 0 & 0 & \frac{3 P}{2} & 0 & - a_{zpx} & 0\\ 
- a_{zpx} & 0 & - a_{zmx} & 0 & 0 & 0 & - a_{zmx} & 0 & - a_{zpx}\\ 
0 & - a_{zpx} & 0 & \frac{3 P}{2} & 0 & 0 & 0 & - a_{zmx} & 0\\ 
0 & 0 & 0 & 0 & - a_{zmx} & 0 & 0 & 0 & \frac{3 P}{2}\\ 
0 & 0 & 0 & - a_{zpx} & 0 & - a_{zmx} & 0 & 0 & 0\\ 
0 & 0 & 0 & 0 & - a_{zpx} & 0 & \frac{3 P}{2} & 0 & 0\end{matrix}\right],
\end{equation}
where ${a_{zmx}\equiv\frac{A_{zz}-A_{xx}}{2}}$, ${a_{zpx}\equiv\frac{A_{zz}+A_{xx}}{2}}$. The wave functions enumerated as 
\par \noindent
$\vline M_S=-1, M_I=-1 \rangle$,~$\vline M_S=-1,~M_I=0 \rangle$, $\vline M_S=-1, M_I=1 \rangle$,
$\vline M_S=0, M_I=-1 \rangle$, ..., $\vline M_S=1, M_I=1 \rangle$.      \par
In the first order of perturbation theory for the operator $\hat{\mathbf{V}}$, the corrections to the energy values are equal to zero 
but the wave functions change as 

\begin{equation}\label{eq_wave1st}
\begin{split}
&\Psi_{1} \approx \vline M_S=-1, M_I=-1 \rangle + \frac{3 P }{4 A_{xx}}\vline M_S=-1, M_I=1 \rangle + \frac{a_{zpx}}{Z_{e}}\vline M_S=0, M_I=0 \rangle, \\ 
&\Psi_{2} \approx \vline M_S=-1, M_I=0 \rangle +  \frac{ a_{zmx}}{Z_{e}}\vline M_S=0, M_I=-1 \rangle + \frac{ a_{zpx}}{Z_{e}}\vline M_S=0, M_I=1 \rangle, \\ 
&\Psi_{3} \approx \vline M_S=-1, M_I=1 \rangle - \frac{3 P }{4 A_{xx}}\vline M_S=-1, M_I=-1 \rangle +  \frac{_{zmx}}{Z_{e}}\vline M_S=0, M_I=0 \rangle, \\ 
&\Psi_{4} \approx  \vline M_S=0, M_I=-1 \rangle +\frac{3 P }{4 Z_n}\vline M_S=0, M_I=1 \rangle- \frac{ a_{zmx}}{Z_{e}}\vline M_S=-1, M_I=0 \rangle +\\
&+ \frac{a_{zpx}}{Z_{e}}\vline M_S=0, M_I=0 \rangle, \\ 
&\Psi_{5} \approx \vline M_S=0, M_I=0 \rangle   - \frac{a_{zpx}}{Z_{e}}\vline M_S=-1, M_I=-1 \rangle - \frac{a_{zmx}}{Z_{e}}\vline M_S=-1, M_I=1 \rangle +\\
&+ \frac{a_{zmx}}{Z_{e}}\vline M_S=0, M_I=-1 \rangle + \frac{a_{zpx}}{Z_{e}}\vline M_S=0, M_I=1 \rangle,\\ 
&\Psi_{6} \approx \vline M_S=0, M_I=1 \rangle - \frac{3 P}{4 Zn}\vline M_S=0, M_I=-1 \rangle- \frac{a_{zpx}}{Z_{e}}\vline M_S=-1, M_I=0 \rangle + \\
&+  \frac{a_{zmx}}{Z_{e}}\vline M_S=0, M_I=0 \rangle, \\ 
&\Psi_{7} \approx \vline M_S=0, M_I=-1 \rangle - \frac{3 P}{4 A_{xx}} \vline M_S=0, M_I=1 \rangle - \frac{ a_{zmx}}{Z_{e}}\vline M_S=0, M_I=0 \rangle, \\ 
&\Psi_{8} \approx \vline M_S=0, M_I=0 \rangle - \frac{ a_{zpx}}{Z_{e}} \vline M_S=0, M_I=-1 \rangle- \frac{ a_{zmx}}{Z_{e}} \vline M_S=0, M_I=1 \rangle,\\ 
&\Psi_{9} \approx \vline M_S=0, M_I=1 \rangle + \frac{3 P }{4 A_{xx}}\vline M_S=0, M_I=-1 \rangle - \frac{ a_{zpx}}{Z_{e}}\vline M_S=0, M_I=0 \rangle,  \\ 
\end{split}
\end{equation}

\noindent where only the largest variables are left in the denominators. As seen, it results in the mixing the wave functions inside 
the nuclear triplet with good efficiency $\frac{3 P }{4 Z_n} \approx $~7\%, $\frac{3 P }{4 A_{xx}} \approx $~0.9\%, 
and the hyperfine interaction mixes the electron levels with less efficiency $\frac{a_{zmx}}{Z_{e}} \approx $~0.04\% or $\frac{a_{zpx}}{Z_{e}} \approx$~0.07\%.

The energy levels in the second order of perturbation theory can be calculated as

\begin{equation}\label{eq_2st}
E_i^{(2)}=E_i^{(0)}+\sum_{j}{\frac{\langle \Psi_i \vline \hat{\mathbf{H}} \vline \Psi_j \rangle ^2}{E_i^{(0)}-E_j^{(0)}}}
\end{equation}

Utilizing the values of $E_i^{(0)}$ obtained from equation~\ref{eq_energy_1st} allows us to express the equation~\ref{eq_2st} as follows

\allowdisplaybreaks
\begin{equation}\label{eq_energy_2st}
\begin{split}
&E_{1}^{(2)}=E_{1}^{(1)} - \frac{a_{zpx}^{2}}{Z_e} +\frac{9P^2}{8 A_{zz}},\\
&E_{2}^{(2)}=E_{2}^{(1)}+- \frac{a_{zmx}^{2}}{Z_e} - \frac{a_{zpx}^{2}}{Z_e},\\
&E_{3}^{(2)}=E_{3}^{(1)} - \frac{a_{zmx}^{2}}{Z_e} -\frac{9P^2}{ 8 A_{zz} },\\
&E_{4}^{(2)}=E_{4}^{(1)} + \frac{a_{zmx}^{2}-a_{zpx}^{2}}{Z_e} +\frac{9P^2}{8 Z_n},\\
&E_{5}^{(2)}=E_{5}^{(1)},\\
&E_{6}^{(2)}=E_{6}^{(1)} - \frac{a_{zmx}^{2}-a_{zpx}^{2}}{Z_e}  - \frac{9P^2}{8 Z_n},\\
&E_{7}^{(2)}=E_{7}^{(1)} + \frac{a_{zmx}^{2}}{Z_e}- \frac{9P^2}{ 8 A_{zz} },\\
&E_{8}^{(2)}=E_{8}^{(1)}+\frac{a_{zmx}^{2}}{Z_e} + \frac{a_{zpx}^{2}}{Z_e},\\
&E_{9}^{(2)}=E_{9}^{(1)} + \frac{a_{zpx}^{2}}{Z_e} +\frac{9P^2}{8 A_{zz} },
\end{split}
\end{equation}
\noindent where only the largest terms are left in the denominators. Then the positions (RF frequencies) of the ENDOR lines can be expressed as

\begin{equation}\label{eq_lines_2st}
\begin{split} 
&f_{L1}^{(2)}=f_{L1}^{(1)}+\frac{\left(A_{xx} - A_{zz}\right)^{2}}{4 Z_e} +\frac{9P^2}{8 A_{zz}},\\
&f_{L2}^{(2)}=f_{L2}^{(1)}- \frac{\left(A_{xx} + A_{zz}\right)^{2}}{4 Z_e} +\frac{9P^2}{8 A_{zz}},\\
&f_{L3}^{(2)}=f_{L3}^{(1)}-\frac{A_{xx} A_{zz}}{Z_e}+\frac{9P^2}{8 Z_n},\\
&f_{L4}^{(2)}=f_{L4}^{(1)}-\frac{A_{xx} A_{zz}}{Z_e}+\frac{9P^2}{8 Z_n},\\
&f_{L5}^{(2)}=f_{L5}^{(1)}+\frac{\left(A_{xx} + A_{zz}\right)^{2}}{4 Z_e} -\frac{9P^2}{8 A_{zz}},\\
&f_{L6}^{(2)}=f_{L6}^{(1)}- \frac{\left(A_{xx} - A_{zz}\right)^{2}}{4 Z_e} -\frac{9P^2}{8 A_{zz}}.
\end{split}
\end{equation}

The theoretical and experimental data merit a comparative analysis. 
The ENDOR spectra facilitate the numerical determination of parameters such as $\frac{A_{xx} A_{zz}}{Z_e}$~=~-0.04~MHz, $\frac{9P^2}{8 A_{xx}}$~=~0.004~MHz, and $\frac{9P^2}{8 Z_n}$~=~0.03~MHz. 
Consequently, for  $\mathbf{B_0} \parallel c$, the calculated line shift is  $-\frac{A_{xx} A_{zz}}{Z_e}+\frac{9P^2}{8 A_{xx}}$~=~0.07~MHz, 
which aligns closely with the experimentally derived value of $0.06\pm0.01$~MHz.  
For $\mathbf{B_0}\perp c$, the model forecasts a range of line shifts between -$\frac{(A_{zz}+A_{xx})^2}{9Z_e}-\frac{(2A_{zz}-A_{xx})^2}{9Z_e}$~=~0.035~MHz   to $-\frac{A_{zz}^2}{Z_e}$~=~0.08~MHz. 
These values are consistent with the experimental data, which spans from $0.03\pm0.01$ and $0.09\pm0.01$~MHz.

\section{Summary}

In this investigation, we thoroughly explored the effects of the hyperfine interaction between 
the electron spin and the three nearest nitrogen atoms ($^{14}$N) on the resulting ENDOR spectra 
of the boron vacancy defect V$_\mathrm{B}^-$ in hBN crystal in the W-band. Measurements of the 
ENDOR spectra for two crystal orientations reveals that this interaction induces a minor but measurable 
in the W-band shift in the effective nuclear g-factor of $^{14}$N, deviating from its established 
tabulated value. This shift is accompanied by a corresponding change in the $^{14}$N Larmor frequency, 
which we interpret using second-order perturbation theory. Considering the impact of one type of 
nuclear sublattice on another is essential when developing algorithms for quantum manipulations 
that involve electronic and nuclear subsystems. This consideration is particularly relevant in 
the context of enhancing the precision of temperature and magnetic field measurements, such as 
those conducted by (nano)sensors utilizing boron defects in hBN.

\acknowledgments
Financial support of the Russian Science Foundation under Grant RSF~24-12-00151~is~acknowledged.

\itemsep=0pt
\bibliography{hBN_ENDOR_secorder_Lit}

\begin{thebibliography}{11}%
\makeatletter
\providecommand \@ifxundefined [1]{%
 \@ifx{#1\undefined}
}%
\providecommand \@ifnum [1]{%
 \ifnum #1\expandafter \@firstoftwo
 \else \expandafter \@secondoftwo
 \fi
}%
\providecommand \@ifx [1]{%
 \ifx #1\expandafter \@firstoftwo
 \else \expandafter \@secondoftwo
 \fi
}%
\providecommand \natexlab [1]{#1}%
\providecommand \enquote  [1]{``#1''}%
\providecommand \bibnamefont  [1]{#1}%
\providecommand \bibfnamefont [1]{#1}%
\providecommand \citenamefont [1]{#1}%
\providecommand \href@noop [0]{\@secondoftwo}%
\providecommand \href [0]{\begingroup \@sanitize@url \@href}%
\providecommand \@href[1]{\@@startlink{#1}\@@href}%
\providecommand \@@href[1]{\endgroup#1\@@endlink}%
\providecommand \@sanitize@url [0]{\catcode `\\12\catcode `\$12\catcode
  `\&12\catcode `\#12\catcode `\^12\catcode `\_12\catcode `\%12\relax}%
\providecommand \@@startlink[1]{}%
\providecommand \@@endlink[0]{}%
\providecommand \url  [0]{\begingroup\@sanitize@url \@url }%
\providecommand \@url [1]{\endgroup\@href {#1}{\urlprefix }}%
\providecommand \urlprefix  [0]{URL }%
\providecommand \Eprint [0]{\href }%
\providecommand \doibase [0]{http://dx.doi.org/}%
\providecommand \selectlanguage [0]{\@gobble}%
\providecommand \bibinfo  [0]{\@secondoftwo}%
\providecommand \bibfield  [0]{\@secondoftwo}%
\providecommand \translation [1]{[#1]}%
\providecommand \BibitemOpen [0]{}%
\providecommand \bibitemStop [0]{}%
\providecommand \bibitemNoStop [0]{.\EOS\space}%
\providecommand \EOS [0]{\spacefactor3000\relax}%
\providecommand \BibitemShut  [1]{\csname bibitem#1\endcsname}%
\let\auto@bib@innerbib\@empty
\bibitem {l4}%
  \BibitemOpen
  \bibfield  {author} {\bibinfo {author} {Gottscholl~A.}, \bibinfo {author}
  {Kianinia~M.}, \bibinfo {author} {Soltamov~V.}, \bibinfo {author}
  {Orlinskii~S.}, \bibinfo {author} {Mamin~G.}, \bibinfo {author} {Bradac~C.},
  \bibinfo {author} {Kasper~C.}, \bibinfo {author} {Krambrock~K.}, \bibinfo
  {author} {Sperlich~A.}, \bibinfo {author} {Toth~M.}, \bibinfo {author}
  {Aharonovich~I.}, \bibinfo {author} {Dyakonov~V.},\ }\href {\doibase
  10.1038/s41563-020-0619-6} {\bibfield  {journal} {\bibinfo  {journal} {\emph
  {Nature Materials}}\ }\textbf {\bibinfo {volume} {19}},\ \bibinfo {pages}
  {540} (\bibinfo {year} {2020})}\BibitemShut {NoStop}%
\bibitem {l2}%
  \BibitemOpen
  \bibfield  {author} {\bibinfo {author} {Astakhov~G.}, \bibinfo {author}
  {Simin~D.}, \bibinfo {author} {Dyakonov~V.}, \bibinfo {author} {Yavkin~B.},
  \bibinfo {author} {Orlinskii~S.}, \bibinfo {author} {Proskuryakov~I.},
  \bibinfo {author} {Anisimov~A.}, \bibinfo {author} {Soltamov~V.}, \bibinfo
  {author} {Baranov~P.},\ }\href {\doibase 10.1007/s00723-016-0800-x}
  {\bibfield  {journal} {\bibinfo  {journal} {\emph {Applied Magnetic
  Resonance}}\ }\textbf {\bibinfo {volume} {47}},\ \bibinfo {pages} {793}
  (\bibinfo {year} {2016})}\BibitemShut {NoStop}%
\bibitem {cobarrubia2024hexagonal}%
  \BibitemOpen
  \bibfield  {author} {\bibinfo {author} {Cobarrubia~A.}, \bibinfo {author}
  {Schottle~N.}, \bibinfo {author} {Suliman~D.}, \bibinfo {author}
  {Gomez-Barron~S.}, \bibinfo {author} {Patino~C.~R.}, \bibinfo {author}
  {Kiefer~B.}, \bibinfo {author} {Behura~S.~K.},\ }\href@noop {} {\bibfield
  {journal} {\bibinfo  {journal} {\emph {ACS nano}}\ }\textbf {\bibinfo
  {volume} {18}},\ \bibinfo {pages} {22609} (\bibinfo {year}
  {2024})}\BibitemShut {NoStop}%
\bibitem {liu2022spin}%
  \BibitemOpen
  \bibfield  {author} {\bibinfo {author} {Liu~W.}, \bibinfo {author}
  {Guo~N.-J.}, \bibinfo {author} {Yu~S.}, \bibinfo {author} {Meng~Y.}, \bibinfo
  {author} {Li~Z.-P.}, \bibinfo {author} {Yang~Y.-Z.}, \bibinfo {author}
  {Wang~Z.-A.}, \bibinfo {author} {Zeng~X.-D.}, \bibinfo {author} {Xie~L.-K.},
  \bibinfo {author} {Li~Q.}, \bibinfo {author} {Wang~J.-F.}, \bibinfo {author}
  {Xu~J.-S.}, \bibinfo {author} {Wang Yi-Tao~J.-S., Tang}, \bibinfo {author}
  {Li~C.-F.}, \bibinfo {author} {Guo~G.-C.~G.},\ }\href@noop {} {\bibfield
  {journal} {\bibinfo  {journal} {\emph {Materials for Quantum Technology}}\
  }\textbf {\bibinfo {volume} {2}},\ \bibinfo {pages} {032002} (\bibinfo {year}
  {2022})}\BibitemShut {NoStop}%
\bibitem {zhang2020material}%
  \BibitemOpen
  \bibfield  {author} {\bibinfo {author} {Zhang~G.}, \bibinfo {author}
  {Cheng~Y.}, \bibinfo {author} {Chou~J.-P.}, \bibinfo {author} {Gali~A.},\
  }\href@noop {} {\bibfield  {journal} {\bibinfo  {journal} {\emph {Applied
  Physics Reviews}}\ }\textbf {\bibinfo {volume} {7}} (\bibinfo {year}
  {2020})}\BibitemShut {NoStop}%
\bibitem {zhang2017depth}%
  \BibitemOpen
  \bibfield  {author} {\bibinfo {author} {Zhang~W.}, \bibinfo {author}
  {Zhang~J.}, \bibinfo {author} {Wang~J.}, \bibinfo {author} {Feng~F.},
  \bibinfo {author} {Lin~S.}, \bibinfo {author} {Lou~L.}, \bibinfo {author}
  {Zhu~W.}, \bibinfo {author} {Wang~G.},\ }\href@noop {} {\bibfield  {journal}
  {\bibinfo  {journal} {\emph {Physical Review B}}\ }\textbf {\bibinfo {volume}
  {96}},\ \bibinfo {pages} {235443} (\bibinfo {year} {2017})}\BibitemShut
  {NoStop}%
\bibitem {yavkin2019epr}%
  \BibitemOpen
  \bibfield  {author} {\bibinfo {author} {Yavkin~B.}, \bibinfo {author}
  {Gafurov~M.}, \bibinfo {author} {Volodin~M.}, \bibinfo {author} {Mamin~G.},
  \bibinfo {author} {Orlinskii~S.~B.},\ }\href@noop {} {\bibfield  {journal}
  {\bibinfo  {journal} {\emph {Experimental Methods in the Physical Sciences}}\
  }\textbf {\bibinfo {volume} {50}},\ \bibinfo {pages} {83} (\bibinfo {year}
  {2019})}\BibitemShut {NoStop}%
\bibitem {murzakhanov2024photoinduced}%
  \BibitemOpen
  \bibfield  {author} {\bibinfo {author} {Murzakhanov~F.}, \bibinfo {author}
  {Latypova~L.}, \bibinfo {author} {Mamin~G.}, \bibinfo {author}
  {Sadovnikova~M.}, \bibinfo {author} {von Bardeleben~H.}, \bibinfo {author}
  {Gafurov~M.},\ }\href@noop {} {\bibfield  {journal} {\bibinfo  {journal}
  {\emph {Magnetic Resonance in Solids}}\ }\textbf {\bibinfo {volume} {26}},\
  \bibinfo {pages} {24208} (\bibinfo {year} {2024})}\BibitemShut {NoStop}%
\bibitem {mackenzie2002multinuclear}%
  \BibitemOpen
  \bibfield  {author} {\bibinfo {author} {MacKenzie~K.~J.}, \bibinfo {author}
  {Smith~M.~E.},\ }\href@noop {} {\emph {\bibinfo {title} {Multinuclear
  solid-state nuclear magnetic resonance of inorganic materials}}},\
  Vol.~\bibinfo {volume} {6}\ (\bibinfo  {publisher} {Elsevier},\ \bibinfo
  {year} {2002})\ p.\ \bibinfo {pages} {23–108}\BibitemShut {NoStop}%
\bibitem {l15}%
  \BibitemOpen
  \bibfield  {author} {\bibinfo {author} {Gracheva~I.~N.}, \bibinfo {author}
  {Murzakhanov~F.~F.}, \bibinfo {author} {Mamin~G.~V.}, \bibinfo {author}
  {Sadovnikova~M.~A.}, \bibinfo {author} {Gabbasov~B.~F.}, \bibinfo {author}
  {Mokhov~E.~N.}, \bibinfo {author} {Gafurov~M.~R.},\ }\href {\doibase
  10.1021/acs.jpcc.2c08716} {\bibfield  {journal} {\bibinfo  {journal} {\emph
  {Journal of Physical Chemistry C}}\ }\textbf {\bibinfo {volume} {127}},\
  \bibinfo {pages} {3634} (\bibinfo {year} {2023})}\BibitemShut {NoStop}%
\bibitem {l12}%
  \BibitemOpen
  \bibfield  {author} {\bibinfo {author} {Murzakhanov~F.~F.}, \bibinfo {author}
  {Mamin~G.~V.}, \bibinfo {author} {Orlinskii~S.~B.}, \bibinfo {author}
  {Gerstmann~U.}, \bibinfo {author} {Schmidt~W.~G.}, \bibinfo {author}
  {Biktagirov~T.}, \bibinfo {author} {Aharonovich~I.}, \bibinfo {author}
  {Gottscholl~A.}, \bibinfo {author} {Sperlich~A.}, \bibinfo {author}
  {Dyakonov~V.}, \bibinfo {author} {Soltamov~V.~A.},\ }\href {\doibase
  10.1021/acs.nanolett.1c04610} {\bibfield  {journal} {\bibinfo  {journal}
  {\emph {Nano Letters}}\ }\textbf {\bibinfo {volume} {22}},\ \bibinfo {pages}
  {2718} (\bibinfo {year} {2022})}\BibitemShut {NoStop}%
\end{thebibliography}%


%
%
%
%

\end{document}